\documentclass[11pt]{article}
\usepackage{amsmath}

\usepackage{relsize}
\usepackage{color}
\title{{\bf ELECTROMAGNETIC SELF-FORCE OF A POINT CHARGE FROM THE RATE OF CHANGE OF THE MOMENTUM OF ITS RETARDED SELF-FIELD}}
\author{V. HNIZDO\thanks{2044 Georgian Lane, Morgantown, WV 26508, USA } \;and G. VAMAN\thanks{Institute of Atomic Physics, P. O. Box MG-6, Bucharest, Romania (vaman@ifin.nipne.ro)}}
\date{}
\begin{document}
\maketitle
\begin{abstract}
\noindent
The self-force of a point charge moving on a rectilinear trajectory is obtained, with no need of any explicit removal of infinities, as the negative of the time  rate of change of the momentum of its retarded self-field.
\end{abstract}
\section*{1. Introduction}
The problem of the self-force, or radiation reaction, of a non-uniformly moving charged particle has a long history. The Abraham-Lorentz evaluation of the self-force \cite{lore}, dating to the early 1900's, assumed the particle's charge distribution to be spherically symmetric, of radius $a$, and obtained the self-force as an expansion in powers of $a$ (including the power $a^{-1}$). Dirac \cite{dir} was the first to obtain the radiation-reaction force assuming at the outset the particle to be point-like, in a relativistic calculation that used conservation of energy and momentum, and, unlike the Abraham-Lorentz evaluation, both the retarded and advanced self-fields of the charge. For a recent survey of the history of this topic see Ref. \cite{mcd}.

In this paper, we present a novel calculation of the self-force of a point charge, using only its retarded self-field. As in the account of Jackson of the Abraham-Lorentz evaluation (\cite{jac}, Sec. 16.3), our starting points are (i) the conservation of momentum in the system consisting of a single charged particle and an external electromagnetic field,
\begin{align}\label{ecc1}
\frac{d \vec{p}_{\rm{mech}}}{dt}+ \frac{d \vec{G}}{dt}=0,
\end{align}
where $\vec{p}_{\rm{mech}}$ is the purely mechanical, or ``bare'', momentum of the particle, and $\vec{G}$ is the momentum of the total electromagnetic field (external plus the particle's self-field), and (ii) the relation
\begin{align}\label{ecc2}
\frac{d \vec{p}_{\rm{mech}}}{dt}= \vec{F}_{\rm{ext}}+ \vec{F}_{\rm{s}}.
\end{align}
Here, $\vec{F}_{\rm{ext}}$ and $\vec{F}_s$ are the Lorentz forces on the charged particle due to the external field and the particle's self-field, respectively (in the original calculations of Abraham and Lorentz, the particle's momentum was taken to be entirely electromagnetic, entailing that $\vec{p}_{\rm{mech}}=0$, and so the total force $\vec{F}_{\rm{ext}}+ \vec{F}_s$ was assumed to vanish). Equations (\ref{ecc1}) and (\ref{ecc2}) can be combined to write
\begin{align}\label{ecc3}
-\frac{d\vec{G}}{dt}= - \frac{d}{dt}(\vec{G}-\vec{G_s}) - \frac{d\vec{G}_s}{dt}= \vec{F}_{\rm{ext}}+ \vec{F}_{\rm{s}},
\end{align}
where $\vec{G}_s$ is the electromagnetic momentum of the self-field only. Equation (\ref{ecc3}) suggests expressing the self-force $\vec{F}_{\rm{s}}$ and the external force $\vec{F}_{\rm{ext}}$ separately as
\begin{align}
&\vec{F}_{\rm{s}}= - \frac{d\vec{G}_{\rm{s}}}{dt},\\
&\vec{F}_{\rm{ext}}=- \frac{d}{dt}(\vec{G}-\vec{G}_{\rm{s}}).
\end{align}
We shall calculate the self-force $\vec{F}_{\rm{s}}$ using the {\it ansatz} (4), assuming for simplicity that the particle moves on a rectilinear trajectory. Remarkably, the calculation will turn out to yield the exact space part of the radiation-reaction 4-force of Dirac's equation \cite{dir} (now usually called the Lorentz-Abraham-Dirac (LAD) equation to reflect fully its origins), with no need of renormalization or any other explicit removal of infinities.

\section*{2. Calculation of the self-force}
\subsection*{2.1 Retarded self-fields}
We consider a point particle of charge $e$, moving on a rectilinear trajectory $w(t)$, say along the $x$-axis. Its charge and current densities are then given by
\begin{align}\label{ec1}
 \rho(\vec{r},t)=e \delta (\vec{r} -w(t) \hat{\vec{x}}), \; \vec{j}(\vec{r},t)=  \rho(\vec{r},t) \dot{w}(t) \hat{\vec{x}},  
\end{align}
where overdot denotes differentiation with respect to time, and $\delta$ is the Dirac delta function. We assume that the charge density can be expanded as the Taylor series:
\begin{align}\label{ec2}
\rho(\vec{r},t)= e\sum_{n=0}^{\infty} \frac{(-1)^n}{n!} w^n(t) \partial_x^n \delta(\vec{r}),
\end{align}
where $ \partial_x^n$ denotes the partial differentiation with respect to $x$ of order $n$; $w^n(t)
\equiv \left(w(t)\right)^n$. 
The Taylor-series expansion suggests that we consider only small oscillations about an equilibrium position, but our final results will not depend on any amplitude of $w(t)$.

Using Eqs.\ (\ref{ec1}) and (\ref{ec2}), the retarded scalar and vector potentials of the moving charge are calculated to be:
\begin{align}\label{ec3}
\phi( \vec{r}, t)&= \int d\vec{r}\,' \; \frac{ \rho( \vec{r}\,', t- |\vec{r}-  \vec{r}\,'|/c)}{ |\vec{r}-  \vec{r}\,'|}\nonumber \\
&= e \sum_{n=0}^{\infty} \frac{(-1)^n}{n!} \partial_x^n \frac{w^n(t-r/c)}{r},
\end{align}
\begin{align}\label{ec4}
\vec{A}( \vec{r}, t)&= \frac{1}{c} \int d\vec{r}\,' \; \frac{ \vec{j}(  \vec{r}\,', t- |\vec{r}-  \vec{r}\,'|/c)}{ |\vec{r}-  \vec{r}\,'|}\nonumber \\
&= \frac{e}{c} \sum_{n=0}^{\infty} \frac{(-1)^n}{n!} \partial_x^n \frac{ w^n(t-r/c)\dot{w}(t-r/c)}{r} \hat{\vec{x}}.
\end{align}
Here, we used the facts that $\partial_{x'}^n \delta(\vec{r}\,') = \delta^{(n)}(x') \delta(y')\delta(z')$ and 
\begin{equation}
\int_{-\infty}^{\infty} dx' f(\vec{r}-  \vec{r}\,') \delta^{(n)}(x') = (-1)^n \partial_{x'}^n f(\vec{r}- \vec{r}\,')|_{x'=0}= \partial_x^n f(\vec{r}-  \vec{r}\,')|_{x'=0}.
\end{equation}
The $y$- and $z$-components of the retarded electric and magnetic self-fields of the charge,
\begin{align}
\vec{E}(\vec{r},t)= - \nabla \phi(\vec{r},t) - \frac{1}{c}\frac{\partial}{\partial t} \vec{A}(\vec{r},t), \; \; \vec{B}(\vec{r},t)= \nabla \times \vec{A}(\vec{r},t),
\end{align}
are obtained using Eqs.\ (\ref{ec3}) and (\ref{ec4}) to be
\begin{align}\label{ec7}
&E_{y,z}(\vec{r},t) = e \sum_{n=0}^{\infty} \frac{(-1)^{n+1}}{n!} \partial_{y,z} \partial_x^n \frac{w^n(t-r/c)}{r},\\
&B_{y,z}(\vec{r},t) =  \frac{e}{c} \sum_{n=0}^{\infty} \frac{(-1)^n}{n!} \tilde{\partial}_{z,y}\partial_x^n \frac{w^n(t-r/c) \dot{w}(t-r/c)}{r},
\end{align}
where $\tilde{\partial_z}=\partial_z$ and $\tilde{\partial_y}=-\partial_y$. 
\subsection*{2.2 Self-field momentum}
The self-fields create an electromagnetic-self-field momentum
\begin{align}\label{ec9}
\vec{G}_s(t)= &\frac{1}{4\pi c} \int d \vec{r} \;\vec{E}(\vec{r},t) \times \vec{B}(\vec{r},t) \nonumber \\
=& \frac{1}{4\pi c} \int d \vec{r}\;
 \left[ E_y(\vec{r},t) B_z(\vec{r},t)- E_z(\vec{r},t) B_y(\vec{r},t) \right] \hat{\vec{x}};
\end{align}
the $y$- and $z$-components of $\vec{G}_s(t) $ can be seen to vanish already on account of symmetry (the particle is moving only along the $x$-axis).

We evaluate the integral in (\ref{ec9}) in the momentum space:
\begin{align}\label{ec10}
\vec{G}_s(t)= \frac{1}{4\pi c} \frac{1}{8\pi^3} \int d \vec{k} \left[ E_y(\vec{k},t) B_z(-\vec{k},t) - 
E_z(\vec{k},t) B_y(-\vec{k},t) \right] \hat{\vec{x}},
\end{align}
where the Fourier transforms are defined by
\begin{equation}\label{ec11}
f(\vec{k},t) = \int d \vec{r} f(\vec{r},t) e^{i \vec{k} \cdot \vec{r}}.
\end{equation}
Using Eqs.\ (\ref{ec7}) and (\ref{ec11}), we express the Fourier transforms $E_{y,z}(\vec{k},t)$ as
\begin{align}\label{ec12}
E_{y,z}(\vec{k},t)&= e\sum_{n=0}^{\infty} \frac{1}{n!} \int d \vec{r} \left( \partial_{y,z} \partial_x^n e^{i \vec{k} \cdot \vec{r}} \right) \frac{w^n(t-r/c)}{r} \nonumber \\
&=e \sum_{n=0}^{\infty} \frac{i^{n+1}}{n!} k_{y,z} k_x^n \int_0^{\infty} dr \; r \;  w^n(t-r/c) \int d \Omega \;e^{ikr\cos \theta} \nonumber \\
& =4\pi e \sum_{n=0}^{\infty} \frac{i^{n+1}}{n!} \frac{k_{y,z} k_x^n}{k} \int_0^{\infty} dr \sin (kr) w^n(t-r/c),
\end{align}
where, in the $1^{\rm{st}}$ line, we integrated by parts; $k_{x,y,z}$ are the Cartesian components of $\vec{k}$ in the momentum space and $k=| \vec{k}|$. Similar calculation yields, using Eq. (13):
\begin{align}\label{ec13}
B_{y,z}(-\vec{k},t)=  \frac{4\pi e}{c} \sum_{n=0}^{\infty} \frac{(-i)^{n+1}}{n!} \frac{\tilde{k}_{z,y} k_x^n}{k} \int_0^{\infty} dr  \sin(kr) w^n(t-r/c) \dot{w}(t-r/c),
\end{align}
where $\tilde{k}_z=-k_z$ and $\tilde{k}_y=k_y$.
Using Eqs.\ (\ref{ec12}) and (\ref{ec13}), the self-field momentum (\ref{ec10}) is now calculated as follows:
\begin{align}\label{ec14}
\vec{G}_s(t) = &\frac{e^2}{2 \pi^2 c^2} \sum_{n=0}^{\infty} \sum_{p=0}^{\infty} \frac{(-1)^{p+1} i^{n+p+2}}{n!p!} \int_0^{\infty} dk \int d \Omega_{\vec{k}} \; (k_y^2 + k_z^2 )k_x^{n+p} \nonumber \\
& \times \int_0^{\infty} dr \sin(kr) w^n(t-r/c) \int_0^{\infty} dr' \sin(kr') w^p(t-r'/c) \dot{w}(t-r'/c) \hat{\vec{x}} \nonumber \\
= &\frac{4e^2}{\pi c^2} \sum_{\kappa =0}^{\infty} \sum_{n=0}^{2\kappa}  \frac{(-1)^{n+\kappa}}{n! (2\kappa -n)! (2\kappa+1)(2\kappa+3)}   \int_0^{\infty} dk \; k^{2\kappa+2}  \nonumber \\
&\times \int_0^{\infty} dr \int_0^{\infty} dr' \sin(kr) \sin(kr') w^n(t-r/c) w^{2\kappa-n}(t-r'/c) \dot{w}(t-r'/c) \hat{\vec{x}} \nonumber \\
 = &\frac{2e^2}{c} \sum_{\kappa =0}^{\infty} \sum_{n=0}^{2\kappa}\frac{(-1)^{n+1}}{n! (2\kappa-n+1)! (2\kappa+1)(2\kappa+3) c^{2\kappa+3}} \nonumber \\
&\times \int_0^{\infty} dr \; w^n(t-r/c) \frac{d^{2\kappa+3}}{dt^{2\kappa+3}} w^{2\kappa -n+1}(t-r/c) \hat{\vec{x}}.
\end{align}
Here, in the $2^{\rm{nd}}$ equality,  we used the result 
\begin{align}
\int d \Omega_{\vec{k}} k_{y,z}^2 k_x^m = \left\lbrace \begin{array}{l}
                                                          4 \pi k^{m+2}/(m+1)(m+3) , \; m \; \mbox{even}\\
                                                          0 , \; m \; \mbox{odd}
                                                          \end{array}   \right. ,
\end{align}
and changed the summation indices using $n+p=2\kappa$ in view of the fact that only the terms with $n+p$  even contribute to $\vec{G}_s(t)$; in the
last equality, we used the generalized-function identity\footnote{Relation (21) can be obtained from $\int_0^{\infty} dk \;k^m \cos(kx)=\pi i^m \delta^{(m)}(x)$, $m$ even \big(see Ref.\ \cite{ligh},  p.\ 43, Table 1; note that the Fourier transform of $f(x)$ is there defined as $g(y)=\int_{-\infty}^{\infty} dx \; f(x)\;\mbox{exp}(-2\pi i y x)$,  and $\sin(kx) \sin(ky) = \frac{1}{2}\cos(k(x-y))-  \frac{1}{2}\cos(k(x+y))\big).  $}
\begin{align}\label{ec16}
\int_0^{\infty}dk\; k^m \sin(kx) \sin(ky)  =\frac{\pi}{2} i^m \left[ \delta^{(m)}(x-y) -  \delta^{(m)}(x+y) \right],\; m \;\; \rm{even},
\end{align}
where only the first term contributed in the radial integration,
and then the identity
\begin{align}\label{eq17}
\frac{d^m f(t-r/c)}{dr^m} = \left( - \frac{1}{c} \right)^m \frac{d^m f(t-r/c)}{dt^m}.
\end{align}
Transforming the integration variable $r$ to $t'=t-r/c$, the self-field momentum (\ref{ec14}) can now be expressed as
\begin{align}\label{ec18}
\vec{G}_s(t) = - \int_{-\infty}^t dt' F_s(t') \hat{\vec{x}},
\end{align}
where $F_s(t')$ (the subscript anticipates relation (4)) is a force given by 
\begin{equation}\label{aa}
F_s(t)=2e^2 \sum_{k=0}^{\infty} \frac{(2k+2)c^{-(2k+3)}}{(2k+1)(2k+3)!} \sum_{n=0}^{2k}  {{2k+1}\choose{n}}  \left( -w(t)  \right)^n
                           \frac{d^{2k+3}w^{2k-n+1}(t)}{dt^{2k+3}};
\end{equation}
$ {{2k+1}\choose{n}}=(2k+1)!/(2k+1-n)!n!$ is a binomial coefficient.

\subsection*{2.3 Self-force}
We now proceed to evaluate the force (\ref{aa}) in closed form.	
	Using the identity
\begin{equation}
\frac{df(t)}{dt} =-c \frac{df(t-r/c)}{dr}\big{|}_{r=0},
\end{equation}
we can write (\ref{aa}) as
\begin{equation}\label{bb}
F_s(t)=2e^2 \sum_{k=0}^{\infty} \frac{(-1)^{2k+3}(2k+2)}{(2k+1)(2k+3)!} \sum_{n=0}^{2k}{{2k+1}\choose{n}} \left( -w(t)  \right)^n
\frac{d^{2k+3}w^{2k-n+1}(t-r/c)}{dr^{2k+3}}\big{|}_{r=0}.
\end{equation}	
But
\begin{align}\label{cc}
&\sum_{n=0}^{2k}{{2k+1}\choose{n}} \left( -w(t)  \right)^n w^{2k-n+1}(t-r/c)\nonumber\\
& \; \;\;=w^{2k+1}(t-r/c) \sum_{n=0}^{2k}{{2k+1}\choose{n}} \left(- \frac{w(t)}{w(t-r/c)}  \right)^n \nonumber\\
&\; \; \;=w^{2k+1}(t-r/c) \left[ \left( 1-\frac{w(t)}{w(t-r/c)} \right)^{2k+1}- \left( \frac{-w(t)}{w(t-r/c)} \right)^{2k+1}  \right] \nonumber\\
&\; \;\;=\left(w(t-r/c)-w(t)  \right)^{2k+1}+w^{2k+1}(t),
\end{align}
where we used in the $3^{\rm{rd}}$ line the binomial theorem. Using now (\ref{cc}) in (\ref{bb}), we get
\begin{equation}\label{dd}
F_s(t)=2e^2 \sum_{k=0}^{\infty}\frac{(-1)^{2k+3}(2k+2)}{(2k+1)(2k+3)!} \frac{d^{2k+3} \left( w(t-r/c)-w(t) \right)^{2k+1}}{dr^{2k+3}}\big{|}_{r=0},
\end{equation}
since $d^{2k+3} w^{2k+1}(t)/dr^{2k+3}=0$ for all $k \geq 0$. We can use here for the higher-order derivatives the formula
\begin{align} \frac{d^m}{dx^m} f^k(x)= \sum_{j_1+ \dots +j_k=m} {{m}\choose{j_1, \dots, j_k}} f^{(j_1)}(x) \dots f^{(j_k)}(x), \end{align}
 which is the differentiation analogue of the multinomial theorem. This transforms (\ref{dd}) into
\begin{align}\label{ee}
F_s(t)=&2e^2 \sum_{k=0}^{\infty} \frac{(-1)^{2k+3}(2k+2)}{(2k+1)(2k+3)! 
}  \sum_{j_1+ \dots +j_{2k+1}=2k+3} {{2k+3}\choose{j_1, \dots, j_{2k+1}}}  \nonumber \\
&\times \frac{d^{j_1} \left( w(t-r/c) - w(t) \right)}{dr^{j_1}}\big{|}_{r=0} \dots \frac{d^{j_{2k+1}} \left( w(t-r/c) - w(t) \right)}{dr^{j_{2k+1}}}\big{|}_{r=0}.
\end{align}
Utilizing that
\begin{align} \frac{d^j \left( w(t-r/c) -w(t) \right)}{dr^j}\big{|}_{r=0} = \left(1- \delta_{j,0} \right) \left( - \frac{1}{c} \right)^j w^{(j)}(t), \end{align}
Eq.\ (\ref{ee}) can be written more simply as
\begin{equation}\label{ff}
F_s(t) = 2e^2 \sum_{k=0}^{\infty} \frac{2k+2}{(2k+1) c^{2k+3}} \sum_{ \begin{array}{c}
	                                                                j_1+ \dots + j_{2k+1}=2k+3\\
	                                                                j_1, \dots, j_{2k+1} > 0
	                                                                \end{array}}
        \frac{w^{(j_1)}(t)}{j_1!} \dots     \frac{w^{(j_{2k+1})}(t)}{j_{2k+1}!}.            \end{equation}
We note that, because of the summation constraints on the orders $j_i$ of the derivatives $w^{(j_i)}(t)$, only the factors $\dot{w}^{2k}(t) \dddot{w}(t)$ and $ \dot{w}^{2k-1}(t) \ddot{w}^2(t)$ are allowed in Eq. (\ref{ff}), their numbers being $ {{2k+1}\choose{1}} =2k+1$
  and  $ {{2k+1}\choose{2}} =(2k+1)k$, respectively. Equation (\ref{ff}) can thus be still more simplified  to read
  \begin{align}\label{ec28}
  F_s(t)&= 2e^2 \sum_{k=0}^{\infty} \frac{2k+2}{(2k+1) c^{2k+3}} \left[ \frac{2k+1}{3!} \dot{w}^{2k}(t) \dddot{w}(t) + \frac{(2k+1)k}{2! 2!} \ddot{w}^2(t) \dot{w}^{2k-1}(t) \right] \nonumber \\
 &= 2e^2 \sum_{k=0}^{\infty} \frac{k+1}{c^{2k+3}} \left( \frac{1}{3} \dot{w}^{2k}(t) \dddot{w}(t) + \frac{k}{2} \dot{w}^{2k-1}(t) \ddot{w}^2(t) \right). 
  \end{align}
Using now the results $\sum_{k=0}^{\infty} (k+1) x^k =1/(1-x)^2$ and $\sum_{k=0}^{\infty} (k+1)(k+2) x^{2k+1} = 2x/(1-x^2)^3$, $0 \le x <1$, the series in (\ref{ec28}) can be summed, yielding
\begin{align}\label{ec29}
F_s(t) = \frac{2e^2}{3c^3} \gamma^4 \ddot{v}(t) + \frac{2e^2}{c^5} \gamma^6 v(t) \dot{v}^2(t), \; \gamma=(1-v^2(t)/c^2)^{-1/2},
\end{align}
where we now write the velocity $v(t)$ for $\dot{w}(t)$. This is exactly the space part   of the relativistic LAD radiation-reaction force \cite{roh1}
\begin{align}
\vec{F}_{\rm{LAD}}= \frac{2e^2}{3c^3} \gamma^2\left[  \ddot{\vec{v}} + \frac{3\gamma^2}{c^2} (\vec{v} \cdot \dot{\vec{v} })\;\dot{\vec{v}} + \frac{\gamma^2}{c^2} (\vec{v} \cdot \ddot{\vec{v}}) \; \vec{v} + \frac{3\gamma^4}{c^4} (\vec{v} \cdot \dot{\vec{v}})^2 \vec{v} \right],
\end{align}
when that is adapted for a rectilinear motion.
Returning to Eq.\ (\ref{ec18}), we now see that if the limits $t \rightarrow -\infty$ of $\dot{v}(t)$ and $\ddot{v}(t)$ vanish, i.e., if the particle moved uniformly in the remote past, then
\begin{equation}\label{ec30}
-\frac{\vec{G}_s(t)}{dt} = F_s(t) \hat{\vec{x}}.
\end{equation}
In words, the self-force calculated as the negative  of the rate of change of the charge's self-field momentum is the LAD radiation-reaction force for the charge's assumed motion.

\section*{3. Conclusions}
We calculated the time rate of change of the retarded self-field momentum of a point charge moving on a rectilinear trajectory of a remote-past constant velocity, and found that its negative equals the charge's LAD radiation-reaction force. The requisite integration was performed in the momentum space, where, apart from the standard practice of doing the angular integrations first, no renormalization or any other removal of infinities was required (unlike in the traditional treatment, in which the charge has a finite extension $a$ and the self-force contains a term proportional to $1/a$, which diverges in the limit $a \rightarrow 0$). For simplicity, our calculation was performed assuming a rectilinear motion, but we believe that it should be possible to extend it to three-dimensional motion.

There are good reasons, among them the existence of runaway, and pre- and post-acceleration solutions of the LAD equation, to regard the very concept of a point charge as unphysical in classical electrodynamics. The equations of motion that admit no unphysical solutions, like the Landau-Lifshitz, Eliezer and Ford-O'Connell equations, assume explicitly or implicitly the charge to have a finite spatial extension of the order of the classical radius corresponding to its mass \cite{ste}. Be it as it may, our result  supports the commonly held notion of the LAD equation as the ``exact'' equation of motion of a point charge.


\begin{thebibliography}{30}
 
 

\bibitem{lore} H. A. Lorentz:   {\it The Theory of Electrons}, $2^{\rm{nd}}$ ed., Stechert, New York 1916.


\bibitem{dir}P. A. M. Dirac:   Classical theory of radiating electrons, {\it Proc. R. Soc. A} {\bf 167} (1938), 148. 


\bibitem{mcd}K. T. McDonald: On the History of the Radiation Reaction, http://www.physics.princeton.edu/$\sim$mcdonald/examples/selfforce.pdf.

\bibitem{jac}J. D. Jackson: {\it Classical Electrodynamics}, $3^{\rm{rd}}$ ed., John Wiley, New York 1999.

\bibitem{ligh}M. J. Lighthill:   {\it An Introduction to Fourier Analysis and Generalized Functions}, Cambridge University Press, Cambridge 1958.


\bibitem{roh1}F.  Rohrlich:   The dynamics of a charged sphere and the electron, {\it Am. J. Phys.} {\bf 65}    (1997), 1051. 

\bibitem{ste}A. M. Steane: Reduced-order Abraham-Lorentz-Dirac equation and the consistency of classical electromagnetism, {\it Am. J. Phys.} {\bf 83} (2015), 256. 


\end{thebibliography}
\end{document}